\documentclass[%
preprint,
 amsmath,amssymb,
 aps,
pra,
]{revtex4-2}

\usepackage{graphicx}% Include figure files
\usepackage{dcolumn}% Align table columns on decimal point
\usepackage{bm}% bold math
\begin{document}

\preprint{}

\title{$\alpha$ Annealing of Ant Colony Optimization in the infinite-range Ising model}% Force line breaks with \\
%\thanks{}%

\author{Shintaro Mori}
\email{shintaro.mori@hirosaki-u.ac.jp}
\affiliation{
Graduate school of Science and Technology, 
Hirosaki University, \\
Bunkyo-cho 3, Hirosaki, Aomori 036-8561, Japan
}

\author{Taiyo Shimizu}
\email{h24ms110@hirosaki-u.ac.jp}
\affiliation{
Graduate school of Science and Technology, 
Hirosaki University, \\
Bunkyo-cho 3, Hirosaki, Aomori 036-8561, Japan
}

\author{Masato Hisakado}
\email{hisakadom@yahoo.co.jp}
\affiliation{
Nomura Holdings Inc., \\Otemachi 2-2-2, Chiyoda-ku, Tokyo 100-8130, Japan
}

\author{Kazuaki Nakayama}
\email{nakayama@math.shinshu-u.ac.jp}
\affiliation{
Department of Mathematical Sciences,
Faculty of Science, Shinshu University, \\ 
Asahi 3-1-1, Matsumoto, Nagano 390-8621, Japan
}

\date{\today}% It is always \today, today,

\begin{abstract}
Ant colony optimization (ACO) leverages the parameter $\alpha$ to modulate the decision function's sensitivity 
to pheromone levels, balancing the exploration of diverse solutions with the exploitation of promising areas. 
Identifying the optimal value for $\alpha$ and establishing an effective annealing schedule remain significant 
challenges, particularly in complex optimization scenarios. This study investigates the $\alpha$-annealing 
process of the linear Ant System within the infinite-range Ising model to address these challenges. 
Here, "linear" refers to the decision function employed by the ants. By systematically increasing $\alpha$, 
we explore its impact on enhancing the search for the ground state. We derive the Fokker-Planck equation for 
the pheromone ratios and obtain the joint probability density function (PDF) in stationary states. 
As $\alpha$ increases, the joint PDF transitions from a mono-modal to a multi-modal state. 
In the homogeneous fully connected Ising model, $\alpha$-annealing facilitates the transition from a trivial 
solution at $\alpha=0$ to the ground state. The parameter $\alpha$ in the annealing process plays a role analogous 
to the transverse field in quantum annealing. Our findings demonstrate the potential 
of $\alpha$-annealing in navigating complex optimization problems, suggesting its broader application beyond the 
infinite-range Ising model.
\end{abstract}

\maketitle

%\tableofcontents

\section{Introduction}
Ant colony optimization (ACO) is a popular meta-heuristic of swarm intelligence for approximating solutions 
to combinatorial optimization problems \cite{Dorigo:1992,Dorigo:1997}. Inspired by the foraging behavior of ant colonies
\cite{Deneubourg:1987,Pasteels:1987,Pasteels:1999,Camazine:2001,Kirman:1993,Hisakado:2015}, ACO employs simple agents, known as 'ants,' that search 
for the optimal solution through a combination of random search and indirect communication. This stigmergic communication 
involves ants depositing 'pheromone' following the construction and evaluation of a candidate solution, with the pheromone 
quantity reflecting the solution's quality, thereby guiding solution construction. ACO's effectiveness has been demonstrated 
across numerous \textit{NP-hard} combinatorial optimization problems, with its success largely attributable to the cooperative 
interactions among ants via pheromones \cite{Cordon:2002, Dorigo:2010,Li:2022,Tang:2023}.

Following ACO's practical successes, several studies have elucidated its underlying mechanisms. Meuleau and Dorigo illustrated 
the strong relationship between ACO algorithms and stochastic gradient descent, demonstrating that specific ACO forms 
probabilistically converge to a local optimum \cite{Meuleau:2002}. Here, 'convergence' implies ants consistently constructing 
the same solution in ACO. Stützle and Dorigo provided proof of convergence for a class of ACO systems to the globally optimal solution \cite{Stutzle:2002}, 
a finding further substantiated by Gutjahr's proof, which drew parallels with the convergence of simulated 
annealing \cite{Gutjahr:2002,Gutjahr:2003}.

For ACO algorithm performance enhancement, controlling the diversity of candidate solutions is paramount 
\cite{Nakamichi:2001,Randall:2002,Meyer:2008}. Achieving an optimal balance between exploration (solution diversity) and 
exploitation (effective use of available solutions) requires meticulously designed convergence dynamics. Premature convergence 
can restrict exploration to a narrow search space segment, while excessively slow convergence may render the search process 
inefficient.

Meyer has emphasized the critical role of the algorithmic parameter $\alpha$ in controlling diversity 
\cite{Meyer:2004,Meyer:2008,Meyer:2008-2}. $\alpha$ determines how the choice function depends on the pheromone amount $x$, 
represented as $x^{\alpha}$. A low $\alpha$ value encourages ants to explore broadly, while a high $\alpha$ focuses the 
search more narrowly, similar to the role of temperature in simulated annealing. Adjusting $\alpha$ allows for a desirable 
balance between exploration and exploitation. The significance of noise in ACO has also been emphasized 
using stochastic differential equations in both static and dynamic environments \cite{Meyer:2008, Meyer:2008-2,Meyer:2017,Meyer:2017-2}. 
Ants respond to a two-choice question, and the noisy communication among ants prevents them from selecting suboptimal choices.

This paper explores the $\alpha$-annealing process of ACO within the infinite-range Ising model. 
Here, $\alpha$-annealing refers to a systematic method of gradually increasing 
$\alpha$ to balance the trade-off between exploration and exploitation.
We adopt a linear decision function and explore the system through stochastic differential equations (SDEs). 
We derive the stationary solution of the Fokker-Planck equation for the pheromone ratios. 
Our analysis predicts a transition from a mono-modal joint probability density function (PDF) to a multi-modal one 
upon $\alpha$ surpassing a critical threshold ($\alpha_c$). The trajectory of the stationary states induced by changes 
in $\alpha$ bridges the trivial solution and the global minimum of the homogeneous fully connected Ising model. 
The parameter $\alpha$ in the annealing process plays a role analogous to the transverse field in 
quantum annealing \cite{Kadowaki:1998}.

The organization of the paper is as follows: Section \ref{sec:model} introduces our ACO model that searches for the 
ground state of the infinite-range Ising model. We adopt an Ant System (AS) with a linear decision function, 
which is the simplest formulation of the Ant Colony Optimization system.  In Section \ref{sec:PDF}, we derive the SDEs 
for the pheromone ratios and obtain the stationary state of the joint PDF. Section \ref{sec:Ising} studies the transition of 
the PDF for the homogeneous fully connected Ising model. The results are supported by numerical simulation in Section 
\ref{sec:numerical}. The probability of finding the ground state is maximized in the $\alpha$-annealing process. 
Finally, Section \ref{sec:conclusion} summarizes our findings.

\section{\label{sec:model}Linear Ant System and Ground State Search of Ising Model}

We address the problem of identifying the ground state of the Ising model, 
characterized by $N$ binary variables $\{X(i) \in \{0,1\},i=1,\cdots,N\}$\cite{stanley:1987}. 
The system's energy is defined as
\begin{equation}
E[\{X(i)\}]=-\sum_{i}h(i)(2X(i)-1)-\frac{1}{N-1}\sum_{i,j,i\neq j} J(i,j)(2X(i)-1)(2X(j)-1).
\end{equation}
In this model, $J(i,j) \in \mathbb{R}$ signifies the exchange interaction strength, 
and $h(i) \in \mathbb{R}$ represents the external field. 
Without loss of generality, we can assume $J(i,j)=J(j,i)$ and $h(i)\ge 0$. 
The Ising-lattice gas transformation $\sigma(i) = 2X(i) - 1$ maps the binary variables 
$\{X(i) \in \{0,1\}, i=1,\cdots,N\}$ to Ising spin variables $\{\sigma(i) \in \{\pm 1\}, i=1,\cdots,N\}$.
At thermal equilibrium, the joint probability distribution of $\{X(i)\}$ aligns with 
the Boltzmann weight, scaled as $\propto e^{-\beta E[\{X(i)\}]}$, where $\beta$ is the inverse temperature. 
A positive external field ($h(i)>0$) biases towards $X(i)=1$, and a positive exchange 
interaction ($J(i,j)>0$) encourages alignment, i.e., $X(i)=X(j)$.

Considering the homogeneous scenario where $J(i,j) = J$ and $h(i) = h$, the model 
is a homogeneous fully connected Ising model, where all variables interact equally. 
The ground state for $h>0$ is uniformly $X(i)=1$, 
with the energy being $-N(J+h)$. At $h=0$, two ground states exist with the ground 
state energy $-NJ$: $X(i)=1$ for all $i$ and $X(i)=0$ for all $i$. The external 
field breaks the degeneracy and the energy difference between 
these states for $h \neq 0$ is $2Nh$. The energy, given the magnetization 
$m = \sum_{i}(2X(i)-1)/N$, is $-N(hm+Jm^2)$. The energy barrier from 
$m=-1$ to $m=1$ is $N(J-h)$ and makes the ground state discovery ($m=1$) challenging
if $m=-1$ is initially found, especially when $J>>h$ and $h>0$.

In the Ant System (AS) described in this paper, ants sequentially search for the ground state of the Ising model. 
The choice made by the $t$-th ant for $X(i)$ is denoted as $X(i,t) \in \{0,1\}$.
In typical AS implementations, multiple ants search for the optimal solution simultaneously in each iteration. 
However, in this model, only one ant conducts the search during each iteration. 
Since the ants communicate through the pheromones they deposit, this difference is not essential, 
provided that the pheromones do not evaporate too rapidly.
The evaluation of the choice $\{X(i,t)\},i=1,\cdots,N$ is based on the energy value, denoted as $E(t) = 
E[\{X(i,t)\}]$. Ant $t$ deposits pheromones on their choices $\{X(i,t)\}$, 
with the amount of pheromone 
given by the Boltzmann weight $e^{-E(t)}$. 
In our previous work, we studied the case where $h(i) = 1$, $J(i, j) = 0$, 
and the pheromone value was set to $\frac{-E(t) + N}{2}$ \cite{Mori:2024}. Here, the term $N$ in 
$-E(t) + N$ ensures that the pheromone value remains non-negative.
The Boltzmann weight pheromone can avoid 
the negative value of the pheromone, one sees that the approximation in the derivation of SDE 
needs the restrictions $h(i)<<1$ and $J(i,j)<<1$.

We assume that the pheromones evaporate and decrease by a factor of $e^{-1/\tau}$ 
after each iteration, where $\tau$ represents the time scale of the pheromone evaporation.
The total value of pheromones that remains after ant $t$'s choices is,
\begin{equation}
S(t)=\sum_{s=1}^t e^{-E(s)-(t-s)/\tau}.
\label{eq:S}
\end{equation}
The remaining pheromone on the choice $X(k)=x$ is 
\begin{equation}
S_x(k,t)=\sum_{s=1}^{t}e^{-E(s)-(t-s)/\tau}\delta_{X(k,s),x}.
\label{eq:Sx}
\end{equation}
Here, $\delta_{x,y}$ represents the Kronecker delta function, 
which is defined to be 1 if $x=y$ and 0 otherwise. 

Ant $t+1$ makes decisions $\{X(k,t+1),k=1,\cdots,N\}$ based on simple probabilistic rules. 
The information provided by $S_x(k,t)$ gives
ant $t+1$ an indirect clue about the choice $x$. 
In Bayesian statistics, if $S_1(k,t) > S_0(k,t)$, then the posterior probability 
that $X(k,t+1) = 1$ exceeds $\frac{1}{2}$; conversely, it is less than $\frac{1}{2}$ if 
$S_1(k,t) < S_0(k,t)$.
We adopt a linear decision function
with a positive parameter $\alpha$ as follows:
\[
P(X(k,t+1)=1)=(1-\alpha)\frac{1}{2}+\alpha\cdot \frac{S_1(k,t)}{S(t)}
\]
Here, $\alpha$ determines the response of the choice
to the values of the pheromones. $S_1(k,t)=0$ and $S_1(k,t)=S(t)$
are the absorbing states for $\alpha=1$, we restrict $\alpha<1$.
When $\alpha=0$, $P(X(k,t+1)=1)=1/2$ and 
the ants choose at random. As $\alpha$ increases, 
the ants take into account the pheromone in their decisions.
In the typical ACO implementation, the decision function adopts a nonlinear form $
P(X(k,t+1)=x)\propto S_x(k,t)^{\alpha}$.
In the binary choice case, the decision under the case $S_1(k,t)\simeq S(t)/2$
 is crucial. The above linear form approximates the typical decision function in 
 the crucial case ($S_1(k,t)/S(t)\simeq 1/2$) as,
\[
P(X(k,t+1)=1)=\frac{S_1(k,t)^{\alpha}}{S_1(k,t)^{\alpha}+S_0(k,t)^{\alpha}} 
\simeq (1-\alpha)\frac{1}{2}+\alpha \cdot \frac{S_1(k,t)}{S(t)}
\]
 
We denote the ratio of the remaining
pheromones on the choice $X(k)=1$ as $Z(k,t)$,
\begin{equation}
Z(k,t)\equiv \frac{S_1(k,t)}{S(t)} \label{eq:Z}.
\end{equation}
The probability of the choice $X(k,t+1)=1$ is expressed as
\begin{equation}
P(X(k,t+1)=1)=(1-\alpha)\frac{1}{2}+\alpha Z(k,t) \equiv f(Z(k,t)) \label{eq:Px}.
\end{equation}
Here, we introduce a decision function $f(z)$,
\[
f(z)\equiv (1-\alpha)\frac{1}{2}+\alpha z  \label{eq:f}.
\]
The first ant ($t = 1$) makes her choice at random, following 
a Bernoulli distribution for each $k$ from 1 to $N$:
\[
X(k,1) \sim \text{Ber}(1/2), \quad k = 1, \ldots, N.
\]

We denote the history of the process as $H_t$. Here $H_t$ means all
choices $\{X(i,s)\},i=1,\cdots,N,s=1,\cdots,t$.
The conditional expected value of $X(i,t+1)$ under $H_t$ is 
\[
\mathbb{E}[X(k,t+1)|H_t]\equiv E[X(k,t+1)|\{Z(k,t)\}]=f(Z(k,t)).
\]
Likewise, the conditional expected value of $E(t+1)$ under $H_t$ is estimated as,
\begin{eqnarray}
\mathbb{E}[E(t+1)|H_t]&=&-\sum_i h(i)(2f(Z(i,t))-1) \nonumber \\
&-&\frac{1}{N-1}\sum_{i,j,i\neq j}J(i,j)(2f(Z(i,t)-1)(2f(Z(j,t))-1) \nonumber 
\end{eqnarray}
Here, we use the fact that $X(i,t+1)$ and $X(j,t+1)$ are conditionally independent.
We also introduce the conditional expected value of $\sigma(i,t+1)=2X(i,t+1)-1$
under $H_t$, which we call ''magnetization'' $M(i,t)$, as
\[
M(i,t)\equiv \mathbb{E}[2X(k,t+1)-1|H_t]=
2(f(Z(i,t)))-1=2\alpha\left(Z(i,t)-\frac{1}{2}\right)\in [-\alpha,\alpha].
\]
The conditional expected value of $E(t+1)$ is 
expressed with $\{M(i,t)\},i=1,\cdots,N$ as
\[
\mathbb{E}[E(t+1)|H_t]
=-\sum_i h(i)M(i,t)-\frac{1}{N-1}\sum_{i,j,i\neq j}J(i,j)M(i,t)M(j,t).
\]

\section{\label{sec:PDF}Dynamics of Pheromone Ratios and Stationary Distribution}

In this section, we investigate the temporal evolution of the system, 
focusing on the dynamics of pheromone ratios, $Z(k,t)=S_1(k,t)/S(t)$. 
Starting from the recursive 
relationship for $S(t)$,
\begin{equation}
S(t+1)=S(t)e^{-1/\tau}+e^{-E(t+1)} \label{eq:S_t+1},
\end{equation}
we examine $\Delta S(t) = S(t+1) - S(t)$, especially in the regime 
where $\tau >> 1$, leading to
\[
\Delta S(t) \approx -\frac{1}{\tau}S(t) + e^{-E(t+1)}.
\]
This difference equation illustrates the rate of change of $S(t)$ over time. 
In the continuous time limit, the differential equation 
for $S(t)$ is obtained as
\[
dS(t) = \left(-\frac{1}{\tau}S(t) + \mathbb{E}[e^{-E(t+1)}|H_t]\right) dt.
\]
Here, we neglect the random force term from the variance of $e^{-E(t+1)}$.
Assuming the system reaches a stationary state as $t \to \infty$, 
$S(t)$ converges to $\tau \mathbb{E}_{st}[e^{-E(t+1)}]$.
The subscript ${}_{st}$ on $\mathbb{E}_{st}[\,\,]$ signifies that the average is taken in the stationary 
distribution of the process $\{X(i,t)\}$.
The expected value of $S(t)$ in this stationary state, denoted as $S_{st}$, is given by
\[
S_{st} \equiv \tau \mathbb{E}_{st}[e^{-E(t+1)}].
\]

\subsection{Stochastic Differential Equation of Pheromone Ratios}

To derive the SDEs for $\{Z(k,t)\}$, we analyze the temporal evolution of $\{S_x(k,t)\}$. 
Decomposing $S_x(k,t+1)$ provides the foundational step:
\[
S_x(k,t+1) = S_x(k,t)e^{-1/\tau} + e^{-E(t+1)|_{X(k,t+1)=x}}\delta_{X(k,t+1),x}.
\]
We then partition $E(t+1)$ into components based on their dependence on $X(k,t+1)$:
\begin{eqnarray}
E(t+1) &=& -\sum_{i \neq k} h(i)(2X(i,t+1)-1) 
\nonumber \\
&-& \frac{1}{N-1}\sum_{i \neq j, i \neq k, j \neq k}(2X(i,t+1)-1)(2X(j,t+1)-1) \nonumber \\
&-& (2X(k,t+1)-1)\left(h(k)+\frac{1}{N-1}\sum_{l \neq k} 2J(k,l)(2X(l,t+1)-1)\right) \nonumber.
\end{eqnarray}
Introducing the concept of the "effective field" $\hat{h}(k,t+1)$, we define it as follows:
\[
\hat{h}(k,t+1) = h(k) + \frac{1}{N-1}\sum_{l \neq k} 2J(k,l)(2X(l,t+1)-1) = -\frac{1}{2}\frac{\partial E(t+1)}{\partial X(k,t+1)}.
\]
For a choice $X(k,t+1)=x$, the energy $E(t+1)$ simplifies to:
\[
E(t+1)|_{X(k,t+1)=x} = E(t+1) - 2\hat{h}(k)(x - X(k,t+1)).
\]
Assuming a small effective field $\hat{h}(k,t+1)$, which is valid for $h(k) \ll 1$ and $J(k,l) \ll 1$, 
the approximation of $e^{-E(t+1)}|_{X(k,t+1)=x}$ is:
\[
\exp(-E(t+1)|_{X(k,t+1)=x}) \approx \exp(-E(t+1))(1 + 2\hat{h}(k,t+1)(x - X(k,t+1))).
\]
Accordingly, $S_x(k,t+1)$ can be reformulated as:
\[
S_x(k,t+1) = S_x(k,t)e^{-1/\tau} + e^{-E(t+1)}\left(1 + 2\hat{h}(k,t+1)(x - X(k,t+1))\right)\delta_{X(k,t+1),x}.
\]
Leveraging the above formulation and eq. (\ref{eq:S_t+1}), we deduce that:
\begin{eqnarray}
Z(k,t+1) &=& \left(1 - \frac{e^{-E(t+1)}}{S(t+1)}\right) Z(k,t) \nonumber \\
&+& \frac{e^{-E(t+1)}}{S(t+1)} \left(1 + 2\hat{h}(k,t+1)(1 - X(k,t+1))\right) \delta_{X(k,t+1),1} \nonumber.
\end{eqnarray}
The incremental change in $Z(k,t)$ is thus estimated as:
\[
\Delta Z(k,t) \approx \frac{e^{-E(t+1)}}{S(t+1)}\left(\delta_{X(k,t+1),1} - Z(k,t) 
+2\hat{h}(k,t+1)(1 - X(k,t+1))\delta_{X(k,t+1),1}\right).
\]
In the stationary state approximation where $S(t+1)=\tau\mathbb{E}[e^{-E(t+1)}] \approx \tau e^{-E(t+1)}$, we have:
\[
\Delta Z(k,t) \simeq \frac{1}{\tau}\left(\delta_{X(k,t+1),1} - Z(k,t) + 2\hat{h}(k,t+1)(1 - X(k,t+1))\delta_{X(k,t+1),1}\right).
\]
The expected value and the variance of $\Delta Z(k,t)$, 
conditioned on the history $H_t$, are approximated as follows:
\begin{eqnarray}
\mathbb{E}[\Delta Z(k,t)|H_t] &\simeq& \frac{1}{\tau}\left[f(Z(k,t))-Z(k,t) 
\right. \nonumber \\
&+& \left. \mathbb{E}\left[2\hat{h}(k,t+1)|H_t\right](1-f(Z(k,t))) f(Z(k,t)) \right], \nonumber \\
\mathbb{V}[\Delta Z(k,t)|H_t] &\simeq& \frac{1}{\tau^2}\mathbb{V}[\delta_{X(k,t+1),1)}|H_t] = \frac{1}{\tau^2}f(Z(k,t))(1-f(Z(k,t))). \nonumber
\end{eqnarray}
Here, we approximate the expected value of the product of the random variables as
the product of the expected values of the random variables. In addition, we 
neglect the variance of the third term of $\Delta Z(k,t)$, which is 
valid when $\hat{h}(k,t+1) \ll 1$. 

The conditional expected value of the effective field 
$\hat{h}(k,t+1)$ under $H_t$ is,
\begin{eqnarray}
\mathbb{E}[\hat{h}(k,t+1)|H_t]
&=& h(k) + \frac{2}{N-1}\sum_{l \neq k} J(k,l)(2f(Z(l,t))-1) \nonumber \\
&=& h(k) + \frac{2}{N-1}\sum_{l \neq k} J(k,l) 2\alpha\left(Z(l,t)-\frac{1}{2}\right) \nonumber \\
&\equiv& \tilde{h}(k,t) \nonumber.
\end{eqnarray}
We note that the conditional expected value of $\hat{h}(k,t+1)$ under $H_t$ 
is a function of $H_t$.
Given the decision function $f(z) = (1-\alpha)\frac{1}{2} + \alpha(z-\frac{1}{2})$, 
and its complement $1-f(z) = \frac{1}{2} - \alpha(z-\frac{1}{2})$, 
$f(z)(1-f(z)) = \frac{1}{4} - \alpha^2(z-1/2)^2$. We have:
\begin{eqnarray}
&&\mathbb{E}[\Delta Z(k,t)|H_t] \simeq 
\frac{1}{\tau}\left[-(1-\alpha)\left(Z(k,t)-\frac{1}{2}\right)+2\tilde{h}(k,t)
\left(\frac{1}{4} - \alpha^2 \left(Z(k,t)-\frac{1}{2}\right)^2\right) \right], \nonumber \\
&&\mathbb{V}[\Delta Z(k,t)|H_t] \simeq \frac{1}{\tau^2}
\left(\frac{1}{4} - \alpha^2\left(Z(k,t)-\frac{1}{2}\right)^2\right).
\end{eqnarray}

The SDEs describing the dynamics of $\{Z(k,t)\}$ are given by:
\begin{equation}
dZ(k,t)=\mathbb{E}[\Delta Z(k,t)|H_t]dt+\sqrt{\mathbb{V}[\Delta Z(k,t)|H_t]}dW(k,t) \label{eq:SDE}.
\end{equation}
where $\vec{W}(t) = \{W(k,t)\}, k=1,\cdots,N$, represents an independent and identically distributed Wiener process, 
and $d\vec{W}(t)$ follows a $N_{N}(0, I dt)$ distribution.
We denote $d$-dimensional normal distribution with expectation $\vec{\mu}$ and 
variance $\Sigma$ as $N_{d}(\vec{\mu},\Sigma)$.

In multiplying eq. (\ref{eq:SDE}) by $2\alpha$, we obtain the SDEs for 
$\{M(k,t)\}$ as follows:
\begin{eqnarray}
dM(k,t) &=& \frac{1}{\tau}\left(-(1-\alpha)M(k,t)+\alpha \tilde{h}(k,t)(1-M(k,t)^2)\right) dt \nonumber \\
&+& \left(\frac{\alpha}{\tau}\right)\sqrt{1-M(k,t)^2}\,dW(k,t)\, ,\,M(k,t)\in [-\alpha,\alpha] \label{eq:SDE_M}.
\end{eqnarray}
The Fokker-Planck equation for the joint PDF of 
$\vec{M}(t)=(M(1,t),\cdots,M(N,t))$ is \cite{Gardiner:2009},
\begin{eqnarray}
\partial_t p(\vec{m},t) &=& -\frac{1}{\tau} \sum_{k} \partial_{m_k}
\left(-(1-\alpha)m_k
+\alpha \tilde{h}_k(1-m_k^2)\right) p(\vec{m},t) \nonumber \\
&+& \frac{1}{2} \left(\frac{\alpha}{\tau}\right)^2 \sum_{k} \partial^2_{m_k}(1-m_k^2) p(\vec{m},t) \label{eq:FP}, \\
\tilde{h}_k &\equiv & h(k)+\frac{2}{N-1} \sum_{l \neq k} J(k,l) m_l. \nonumber 
\end{eqnarray}

\subsection{Stationary Distribution of Pheromone Ratios}
We derive the stationary solution of the Fokker-Planck equation (\ref{eq:FP}).
We define $A(m|h)$ and $B(m)$ as follows:
\begin{eqnarray}
A(m|h) &=& \frac{1}{\tau}\left(-(1-\alpha)m + \alpha h(1-m^2)\right), \nonumber \\
B(m) &=& \frac{\alpha}{\tau}\sqrt{1-m^2}. \nonumber
\end{eqnarray}
The Fokker-Planck equation (\ref{eq:FP}) can be expressed as:
\[
\partial_t p(\vec{m},t) = \sum_{k} \left\{ -\partial_{m_k} A(m_k|\tilde{h}_k) + \frac{1}{2}\partial^2_{m_k} B^2(m_k) \right\} p(\vec{m},t).
\]
We define $J_k$ as:
\[
J_k \equiv \left\{A(m_k|\tilde{h}_k) - \frac{1}{2} \partial_{m_k} B^2(m_k)\right\} p(\vec{m},t).
\]
Thus, the Fokker-Planck equation simplifies to:
\[
\partial_t p(\vec{m},t) = -\sum_{k} \partial_{m_k} J_k.
\]
To obtain the stationary solution where $\partial_t p(\vec{m}, t) = 0$, we solve for 
$J_k = 0$\cite{Gardiner:2009}. We apply the reflecting boundary condition:
\[
J_k(m_k=\pm \alpha)=0.
\]
From $J_k=0$, we obtain:
\[
\left\{A(m_k|\tilde{h}_k) - \frac{1}{2} \partial_{m_k} B^2(m_k)\right\} p_{st}(\vec{m}) = \left\{\frac{1}{2} B^2(m_k)\right\} \partial_{m_k} p_{st}(\vec{m}).
\]
We define $Z_k$ as:
\[
Z_k \equiv \frac{A(m_k|\tilde{h}_k) - \frac{1}{2} \partial_{m_k} B^2(m_k)}{\frac{1}{2} B^2(m_k)}
= \frac{2 A(m_k|\tilde{h}_k)}{B^2(m_k)} - \partial_{m_k} \log B^2(m_k).
\]
It follows that:
\[
\partial_{m_k} \log p_{st}(\vec{m}) = Z_k.
\]
The potential $\phi(\vec{m})$ for the potential solution $p_{st}(\vec{m})\propto e^{-\phi(\vec{m})}
$ satisfies:
\[
\partial_{m_k} \phi(\vec{m}) = -Z_k.
\]
The existence of $\phi(\vec{m})$ is guaranteed by the condition \cite{Gardiner:2009}: 
\[
\partial_{m_l} Z_k = 2 \left(\frac{\tau}{\alpha}\right) \frac{2}{N-1} J(k,l) = 2 \left(\frac{\tau}{\alpha}\right) \frac{2}{N-1} J(l,k) = \partial_{m_k} Z_l.
\]
The potential $\phi(\vec{m})$ is given by:
\begin{eqnarray}
\phi(\vec{m}) = -\int^{\vec{m}} \vec{Z} d\vec{m}
&=& -\left(\left(\frac{\tau}{\alpha^2}\right) (1-\alpha) - 1\right) \sum_k \log(1-m_k^2) \nonumber \\
&-& 2 \left(\frac{\tau}{\alpha}\right)
\sum_{k} \left\{h(k)+\frac{1}{N-1} \sum_{l \neq k} J(k,l) m_l \right\} m_k. \nonumber 
\end{eqnarray}
The joint PDF of the stationary state $p_{st}(\vec{m})$ is given as:
\begin{eqnarray}
p_{st}(\vec{m}) &\propto& \exp\left(
\sum_{k} \left\{\frac{1}{2} a(\alpha) \log(1-m_k^2) + 2 \left(\frac{\tau}{\alpha}\right)
\left(h(k) m_k + \frac{1}{N-1} \sum_{l \neq k} J(k,l) m_k m_l \right) \right\}
\right), \nonumber \\ 
a(\alpha) &\equiv& 2 \left(\left(\frac{\tau}{\alpha^2}\right) (1-\alpha) - 1\right). 
\end{eqnarray}
We assume the stability of the system and that $\vec{m} = \vec{0}$ should be the 
unique mode for $J(k,l) = h(k) = 0$. We restrict $\alpha$ so that the coefficient $a(\alpha)$ 
of $\log(1-m_k^2)$  is positive. We set the upper bound of $\alpha$
 as $1-\frac{1}{\tau}<1$  and ensure that $a(1-\frac{1}{\tau})=2(1/\alpha^2-1)
> 0$.

In the derivation of the SDEs, we assume that $\tau \gg 1$. 
We neglect the last term $-1$ in $a(\alpha)$, which is valid for $\tau \gg 1$  
and $1-\alpha \gg \frac{1}{\tau}$. We introduce the energy term of the Ising model in the stationary state 
as a function of \( \vec{m} \) as:
\[
E_{\mbox{Ising}}(\vec{m}) = -\sum_{k} h(k) m_k - \frac{1}{N-1} \sum_{k, l, k \neq l}
J(k,l) m_k m_l.
\]
We also introduce the entropy energy of the AS as:
\[
E_{\mbox{AS}}(\vec{m}) = -\sum_{k} \log(1-m_k^2).
\]
The stationary distribution $p_{st}(\vec{m})$ is expressed as:
\begin{equation}
p_{st}(\vec{m}) \propto 
\exp\left(-\left(\frac{2\tau}{\alpha^2}\right)\left[
(1-\alpha) E_{\mbox{AS}}(\vec{m}) + \alpha E_{\mbox{Ising}}(\vec{m}) \right]\right) \label{eq:FE}.
\end{equation}
The terms in the square bracket in the right-hand side of eq. (\ref{eq:FE}) define the "free energy" of the AS. 
When $\alpha \ll 1$, the entropy energy term $E_{\mbox{AS}}(\vec{m})$ dominates the free energy. 
As $m_k \in [-\alpha, \alpha]$, the modes of $p_{st}(\vec{m})$  
should exist near $\vec{0}$. 
A small $\alpha$ initially enables the system to avoid premature convergence by maintaining a broad exploration 
space, which is vital for escaping local minima. 
As $\alpha$ increases, the energy term $E_{\mbox{Ising}}(\vec{m})$ 
begins to dominate the free energy.
The exploration space is restricted to a local minimum of $E_{\mbox{Ising}}(\vec{m})$, 
allowing for intensive exploration and exploitation around the promising regions. 
When $\alpha \approx 1-\frac{1}{\tau} \approx 1$, $\tau \gg 1$, the entropy energy term disappears and the stationary distribution of $\vec{m}$  is governed by the Boltzmann weight $\exp(-(2\tau/\alpha)E_{\mbox{Ising}}(\vec{m}))$.
The inverse temperature $\beta$ of the AS is given by:
\[
\beta = 2\tau/\alpha.
\]
The range of $m_{k} \in [-(1-1/\tau), 1-1/\tau]$ is wide and the mode of $p_{st}(\vec{m})$ 
corresponds to the local minimum of $E_{\mbox{Ising}}(\vec{m})$. 

In terms of Bayesian statistics, the AS provides 
$\exp(-\left(\frac{2\tau}{\alpha^2}\right)(1-\alpha)E_{\mbox{AS}}(\vec{m})), -\alpha \le m_k \le \alpha$ as a prior. 
Multiplied by the likelihood of $\vec{m}$, $\exp(-\beta E_{\mbox{Ising}}(\vec{m}))$, the posterior gives $p_{st}(\vec{m})$.
In order to obtain the global minimum of $E_{\mbox{Ising}}(\vec{m})$ in the $\alpha$-annealing process, 
the inverse temperature $\beta$ should be increased with the increase of $\alpha$.

The essential difference between Simulated Annealing (SA) and $\alpha$-annealing of ACO is the path of the annealing process. 
In $\alpha$-annealing, the system connects the unique and trivial global minimum of the entropy energy $E_{\mbox{AS}}(\vec{m})$ 
and the global minimum of the Ising energy $E_{\mbox{Ising}}(\vec{m})$. This feature reminds us of the similarity between 
$\alpha$-annealing and quantum annealing \cite{Kadowaki:1998}. Additionally, in $\alpha$-annealing, the range of the solution 
$\vec{m}_*$ should be restricted as $m_{*,k} \in [-\alpha, \alpha]$. With these two factors, the $\alpha$-annealing process 
addresses the problem of exploration-exploitation trade-off.

The modes $\vec{m}_{*}$ of $p_{st}(\vec{m})$ satisfy the following relation:
\[
a(\alpha)\frac{m_{k*}}{1-m_{k*}^2} = 2 \left(\frac{\tau}{\alpha}\right)
\left(h(k) + \frac{2}{N-1} \sum_{l \neq k} J(k,l) m_{l*}\right).
\]
This equation corresponds with the TAP equation in spin-glass theory
\cite{TAP:1977}.
The fluctuation of $\vec{m}$ around $\vec{m}_{*}$ can be approximated 
by a Gaussian distribution as:
\[
p_{st}(\Delta \vec{m}) \propto \exp\left(-\frac{1}{2} \Delta \vec{m}^T \Sigma^{-1} \Delta \vec{m}\right),
\]
where \( \Sigma^{-1}_{k,l} \) is given by:
\[
\Sigma^{-1}_{k,l} = \left\{
\begin{array}{cc}
-4 \left(\frac{\tau}{\alpha}\right) \frac{J(k,l)}{N-1} & k \neq l \\
a(\alpha) \frac{(1+m_{k*}^2)}{(1-m_{k*}^2)^2} & k = l
\end{array}
\right.
\]
In the Gaussian approximation, $\vec{m}$ obeys a multi-dimensional normal distribution as:
\[
\vec{m} \sim N_N(\vec{m}_{*}, \Sigma).
\]

In the stationary distribution $p_{st}(\vec{m})$, the local behavior 
around each mode $\vec{m}_{*}$ approximates a normal distribution.
When there are multiple modes, $\{\vec{m}_{*}\}$,
the relative probabilities of the system being near any particular mode are roughly
determined by $p_{st}(\vec{m}_*)$. Consequently, 
the overall distribution of $\vec{m}$ can be characterized as a mixture of 
normal distributions. Each component of this mixture corresponds to a normal 
distribution centered at a mode $\vec{m}_*$, with the mixing weights given by 
the values of $p_{st}(\vec{m}_*)$ at these modes. This formulation captures 
the system’s tendencies towards different stable states under varying conditions, 
reflecting the multimodal nature of the landscape defined by the stationary distribution.

\section{\label{sec:Ising}Homogeneous fully connected Ising model Case}

We study the stationary distribution $p_{st}(\vec{m})$ for
the homogeneous fully connected Ising model. We adopt $J(k,l)=J$ and $h(k)=h$.
The mode $\vec{m}_{*}$ is homogeneous, so we write $\vec{m}_{*}=m_{*}\vec{1}$,
where $\vec{1}$ is an $N$-dimensional vector with all components equal to 1.
We define $b$ as:
\[
b\equiv -4\left(\frac{\tau}{\alpha}\right)\frac{1}{N-1}J.
\]
$m_{*}$ satisfies the following relation:
\begin{equation}
(a(\alpha)+(N-1)b)m_{*}-(N-1)bm_{*}^3-2\left(\frac{\tau}{\alpha}\right)h(1-m_{*}^2)=0.
\label{eq:cubic}
\end{equation}
This is a cubic equation with at most three real solutions.

\begin{figure}[ht]
\begin{center}
\begin{tabular}{cc} 
\includegraphics[width=8cm]{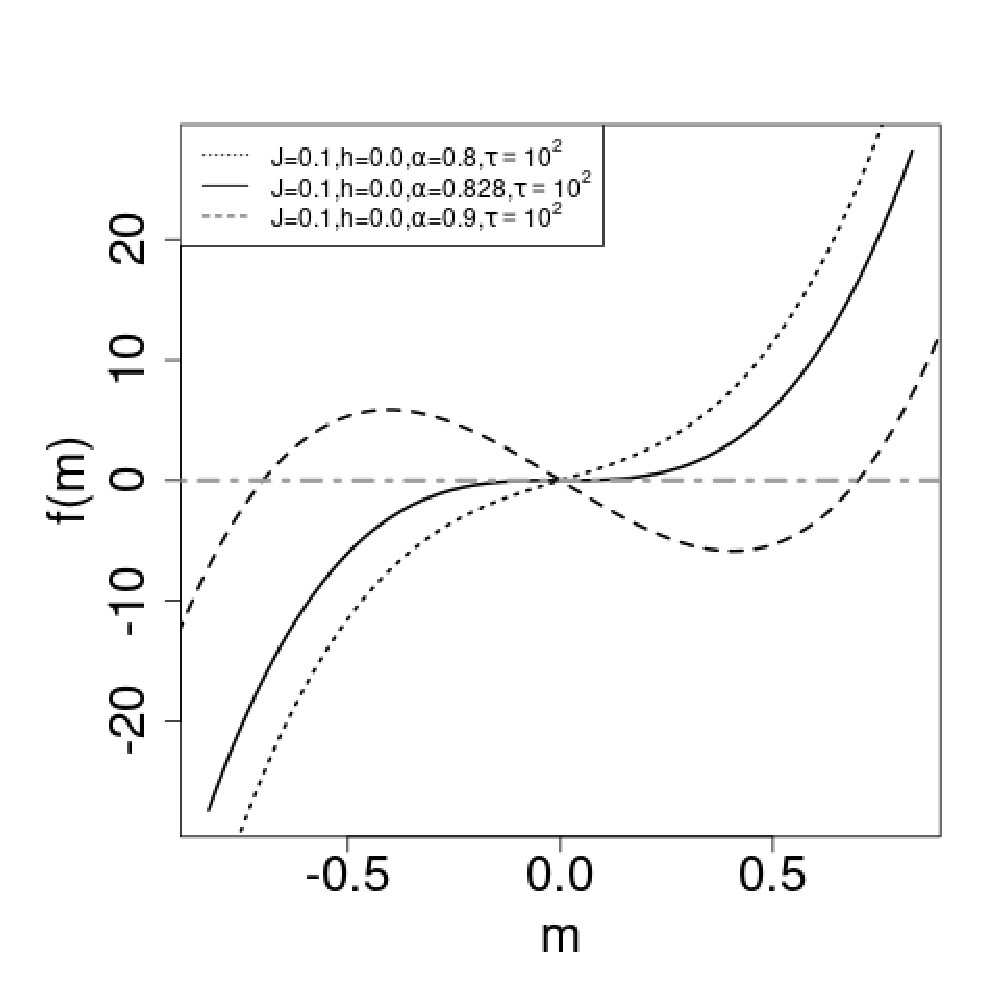}
& 
\includegraphics[width=8cm]{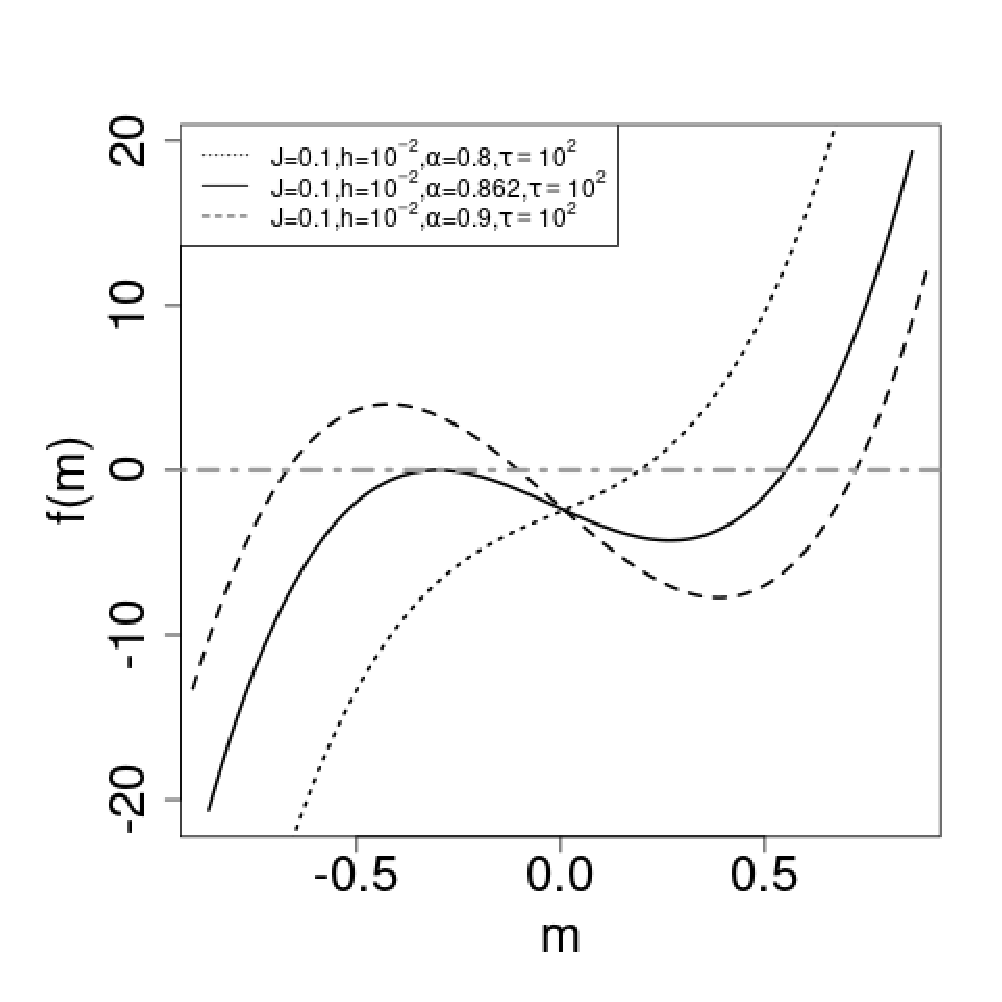}
\end{tabular}
\end{center}
\caption{Plot of cubic equation (\ref{eq:cubic}) vs. $m$.
$J=0.1, h=0.0$ (Left) and 
$J=0.1, h=0.01$ (Right). 
$\alpha=0.8$ (dotted), $\alpha=\alpha_c(h)$ (solid), and $\alpha=0.9$ (broken).
}
\label{fig:1}
\end{figure}	

\subsection{$h=0$ Case}
When $h=0$, $m_{*}$ satisfies:
\begin{equation}
m_{*}((a(\alpha)+(N-1)b)-(N-1)bm_{*}^2)=0 \label{eq:MF}.
\end{equation}
In addition to the solution $m_{*}=0$, when $a(\alpha)+(N-1)b<0$, there appear two 
other real solutions. At $\alpha=\alpha_c$, $a(\alpha_c)+(N-1)b=0$ holds.
$\alpha_c$ is given as:
\[
\alpha_c=\frac{\sqrt{\tau^2(2J+1)^2+4\tau}-\tau(2J+1)}{2} \simeq \frac{1}{1+2J}.
\]
The left figure in Figure \ref{fig:1} shows the plot of the cubic equation (\ref{eq:cubic})
vs. $m$ for $J=0.1, h=0.0$. $\alpha_c=0.82765$, and we choose $\alpha=0.8, \alpha_c$, and $0.9$. 

For $\alpha \le \alpha_c$, $m_{*}=0$ is the unique solution. Above $\alpha_c$,
two other solutions appear: $z_{-}$ and $z_{+}$. They are given as:
\[
z_{+} = -z_{-} = \sqrt{\frac{a(\alpha)+(N-1)b}{(N-1)b}} \propto (\alpha-\alpha_c)^{1/2}.
\]
We summarize the results as:
\[
m_{*}=
\left\{
\begin{array}{cc}
 0  & \alpha \le \alpha_c \\
 0, \pm \sqrt{\frac{a(\alpha)+(N-1)b}{(N-1)b}}   & \alpha > \alpha_c
\end{array}
\right.
\]
For $\alpha \le \alpha_c$, $p_{st}(\vec{m})$ becomes maximal 
at $\vec{m}_{*}=\vec{0}$.
For $\alpha > \alpha_c$,  
$p_{st}(\vec{m})$ becomes maximal at $z_{+}$ and $z_{-}$.
At $\vec{m}=\vec{0}$, $p_{st}(\vec{m})$
becomes minimal.

\subsection{$h>0$ Case}
When $h>0$, there is also a threshold value $\alpha_c(h)$ for $\alpha$. 
For $\alpha < \alpha_c(h)$,
there is a positive real solution, $z_{+}$, where $p_{st}(\vec{m})$ becomes maximal. 
At $\alpha=\alpha_c(h)$, there are two real solutions, $z_{t} < z_{+}$.
The smaller solution $z_{t}$ is a multiple root of eq. (\ref{eq:cubic}), 
and $p_{st}(\vec{m})$
is not maximal. At $z_{+}$, $p_{st}(\vec{m})$ becomes maximal.
For $\alpha > \alpha_c$, there are three real solutions: $z_{-} < z_{u} < z_{+}$. 
We denote the smallest
and the largest solutions as $z_{-}$ and $z_{+}$, respectively.
$p_{st}(\vec{m})$ becomes maximal at these solutions. At the middle solution $z_{u}$,
$p_{st}(\vec{m})$ is minimal.

\subsection{$\alpha_c(h)$ and $m_{*}$}
\begin{figure}[ht]
\begin{center}
\begin{tabular}{cc} 
\includegraphics[width=8cm]{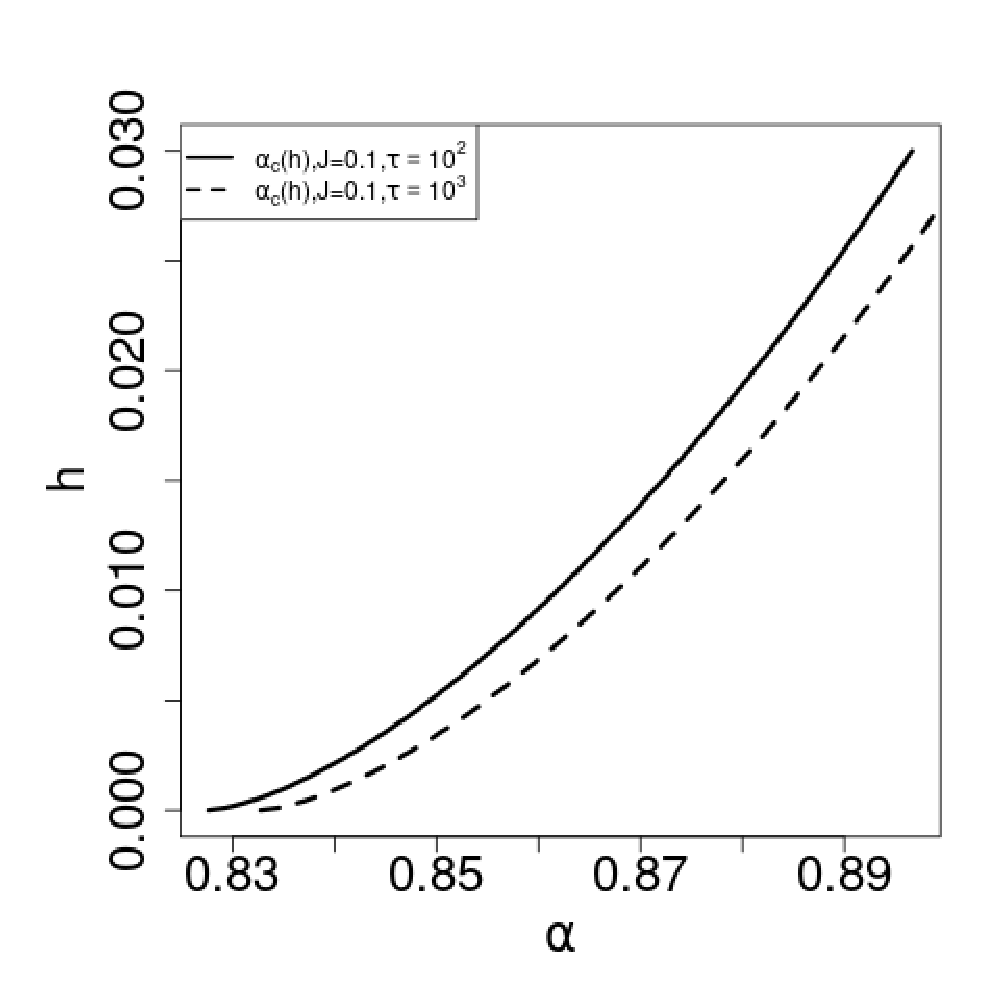}
& 
\includegraphics[width=8cm]{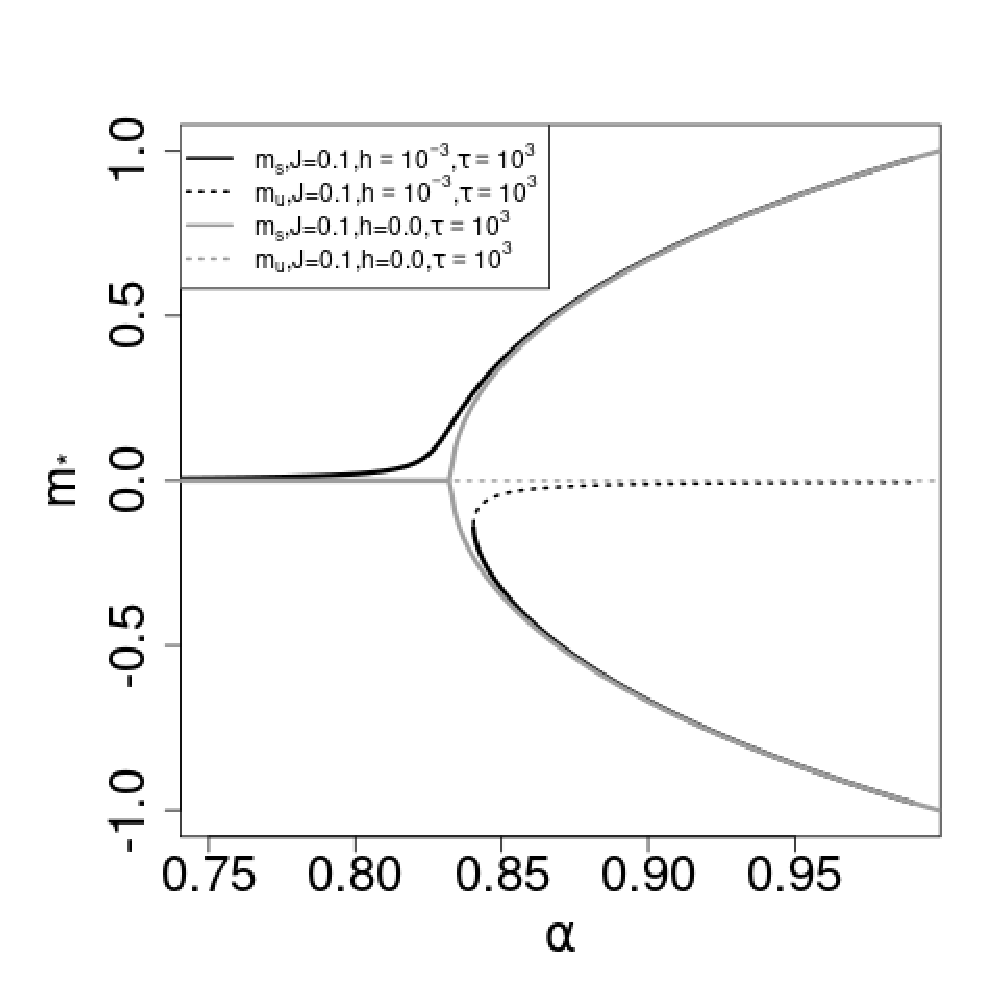}
\end{tabular}
\end{center}
\caption{$h$ vs. $\alpha_c(h)$ (Left)
and 
$m_{*}$ vs. $\alpha$ (Right).
In the left figure, we adopt $J=0.1, \tau=10^2$ (solid line) 
and $J=0.1, \tau=10^3$ (broken line).
In the right figure, we adopt $J=0.1, h=0.0$ (gray) and 
$J=0.1, h=10^{-3}$ (black). The solid lines 
show the solutions where $p_{st}(\vec{m})$ becomes maximal,
and the dotted lines show the 
solutions where $p_{st}(\vec{m})$ becomes minimal.
}
\label{fig:2}
\end{figure}

We solve eq. (\ref{eq:cubic}) numerically to obtain $\alpha_c(h)$.
We also obtain the real solutions $m_{*}$
vs. $\alpha$. Figure \ref{fig:2} summarizes the results.

\subsection{Correlation of $\vec{M}$}
The inverse of the covariance $\Sigma^{-1}$ is given as:
\[
\Sigma^{-1}_{k,l}=
\left\{
\begin{array}{cc}
b & k \neq l \\
a' \equiv a(\alpha) \frac{(1+m_{*}^2)}{(1-m_{*}^2)^2} & k = l \\
\end{array}
\right.
\]
The inverse matrix of $(a'-b)I + bJ$, where $I$ is the identity matrix and $J$ is the matrix 
with all components equal to 1, is given as:
\[
((a'-b)I + bJ)^{-1} = \frac{1}{a'-b}I - \frac{b}{a' + (N-1)b} \cdot \frac{1}{a'-b}J.
\]
Using this result, we obtain $\Sigma$:
\[
\Sigma_{k,l} =
\left\{
\begin{array}{cc}
\frac{1}{a'-b} \cdot \frac{-b}{a' + (N-1)b} & k \neq l \\
\frac{1}{a'-b} \left(1 + \frac{-b}{a' + (N-1)b}\right) & k = l 
\end{array}
\right.
\]
The correlation coefficient between $M_i$ and $M_j$ is:
\[
\mbox{Corr.}(M_i, M_j) = \frac{\Sigma_{i,j}}{\sqrt{\Sigma_{i,i} \Sigma_{j,j}}}
= \frac{-b}{a' + (N-2)b}.
\]
In the case $h=0$, at $\alpha=\alpha_c(h=0)$, $m_{*}=0$ and $a'=a$.
$a(\alpha_c)+(N-1)b=0$ holds, and $\mbox{Corr.}(M_i, M_j)=1$.

\subsection{Marginal pdf of $M_i$}
In the Gaussian approximation, for $\alpha < \alpha_c(h)$,
the marginal distribution of $M_i$ around the mode $m_{*}$ is given as:
\begin{eqnarray}
p_{st}(m_i) &\propto& \exp\left(-\frac{1}{2 \Sigma_{i,i}} (m_i - m_{*})^2 \right) \, , \, m_i \in [-\alpha, \alpha], \nonumber \\
\Sigma_{i,i} &=& \frac{1}{a'-b} \left(1 + \frac{-b}{a' + (N-1)b}\right) \, , \, a' = a(\alpha) \frac{(1 + m_{*}^2)}{(1 - m_{*}^2)^2} \nonumber.
\end{eqnarray}
Here, $m_{*}$ is the unique solution of eq. (\ref{eq:cubic}).

For $h=0$, at the critical point $\alpha=\alpha_c(0)$,
$\Sigma_{i,i}$ diverges and the Gaussian approximation breaks down.
We cannot neglect the higher order terms in $\log(1-m_i^2)$, and $p_{st}(m)$ is given as:
\[
p_{st}(m_i) \propto \exp\left(-\frac{1}{2 \Sigma_{i,i}} (m_i - m_{*})^2 \right)
\cdot \exp\left(\frac{1}{2} a(\alpha) \{\log (1-m_i^2) + m_i^2\} \right).
\]
At the critical point $\alpha=\alpha_c(0)$, the first term on the right-hand side 
of $p_{t}(m_i)$ becomes 1, and the second term describes the PDF.

Above $\alpha_c$, $p_{st}(\vec{m})$ has two modes at $\vec{m}_{+}$ and $\vec{m}_{-}$,
where $\vec{m}_{+}=m_{+}\vec{1}$ and $\vec{m}_{-}=m_{-}\vec{1}$.
We denote the relative probabilities for the two modes $m_{+}$ and $m_{-}$ 
as $p_{+}$ and $p_{-}$, respectively. For $h=0$, $p_{+}=p_{-}=1/2$.
$p_{st}(m_i)$ is the mixture of the two normal distributions approximately:
\[
p_{st}(m_i) \propto 
p_{+} \exp\left(-\frac{1}{2 \Sigma_{i,i}(+)} (m_i - m_{+})^2 \right) +
p_{-} \exp\left(-\frac{1}{2 \Sigma_{i,i}(-)} (m_i - m_{-})^2 \right).
\]
Here, $\Sigma_{i,i}(+)$ and $\Sigma_{i,i}(-)$ are estimated using $m_{+}$ and $m_{-}$, respectively. 
For $h>0$, we need to estimate $p_{+}$ and $p_{-}$ using the relation:
\[
\frac{p_{+}}{p_{-}} = \frac{p_{st}(\vec{m}_+)}{p_{st}(\vec{m}_-)}
\simeq \exp\left(-\frac{2\tau}{\alpha}\{E_{Ising}(\vec{m}_+) - E_{Ising}(\vec{m}_-)\}\right).
\]

\section{\label{sec:numerical}Numerical Study of $\alpha$-annealing}

We have conducted numerical simulations to validate the theoretical predictions 
associated with $\alpha$-annealing in the homogeneous fully connected Ising model.
$\{M(i,t)=2\alpha(Z(i,t)-1/2)\}$ were sampled according to the following annealing schedule:
\[
\alpha(t)= \frac{t}{T}, \quad \alpha(t) < 1-\frac{1}{\tau}, \quad t=0,1,\cdots,
\]
We set $T=10^6$ and $T=10^4$ and refer to them as "slow" and "fast" annealing, respectively.
In the annealing process, the increment of $\alpha$ is given by $\Delta \alpha=1/T$.  
We conducted $S = 1000$ trials for each schedule. 
$M(i, t, s)$ represents the magnetization at time $t$ for $X(i)$ during trial $s$. 

We considered a system size of $N=100$ spins.
We set the parameters as $h = 10^{-3}$, $J = 10^{-1}$, and 
$\tau \in  \{10^{2},10^3\}$. 
In addition, when studying the stationary distribution of $\{M(i,t)\}$ for specific 
$J$, $h$, and $\alpha$,
we adopted the slow annealing schedule with fixed $J$ and $h$.
If $\alpha(t)$ reaches a specific value, we sampled $\{M(i,t)\}$ only once in order to 
ensure the independence of the sampling process.
We repeated the process $5\times 10^3$ times and studied the PDF
of $\{M(i,t,s)\}, s=1,\cdots, 5\times 10^3$. 
The sample size of $\{M(i,t)\}$ is $5\times 10^3 \times N = 5\times 10^5$.

When comparing the performance of $\alpha$-annealing with simulated annealing (SA),
we performed SA with the conventional Metropolis-Hastings update algorithm.
For $\tau$, we set the final inverse temperature as $\beta=2\tau$ and 
the increment of $\beta$ after each Monte Carlo step is set as:
\[
\Delta \beta = \frac{2\tau}{T}.
\]
We have done the sampling process $10^4$ times and estimated 
the success probability to find the ground state of the model.

The conditions for comparison of the two algorithms were kept identical.
In ACO, every ant chose $X(i), i=1,\cdots,N$ and the number of ants was about $10^6$ 
under the slow annealing schedule. While in SA, the final inverse temperature $\beta$ was 
reached after $10^6$ Monte Carlo steps (MCS). In one MCS, the number of trials for the 
spin update is $N$.  

\subsection{Stationary distribution of $M(i,t)$}
\begin{figure}[ht]
\begin{center}
\begin{tabular}{cc} 
\includegraphics[width=8cm]{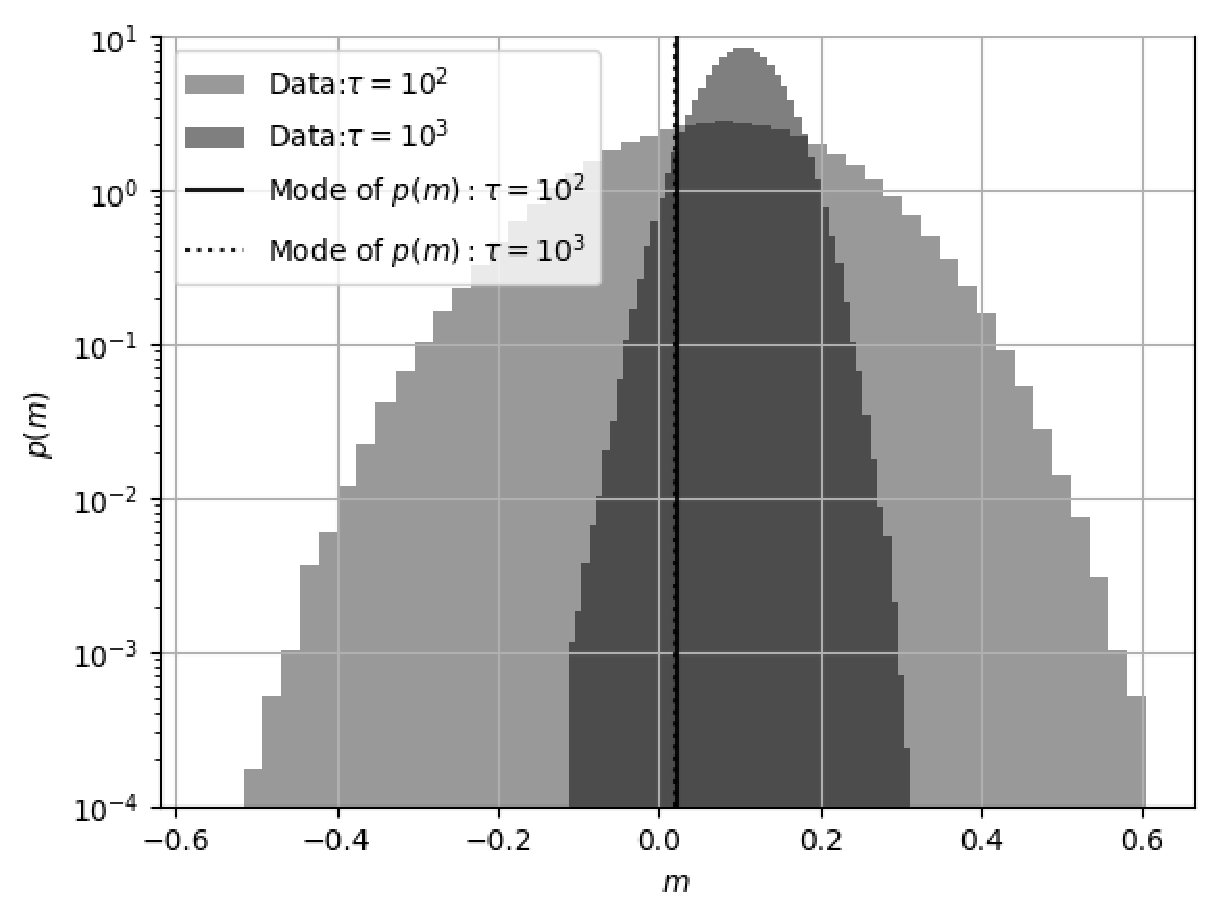}
& 
\includegraphics[width=8cm]{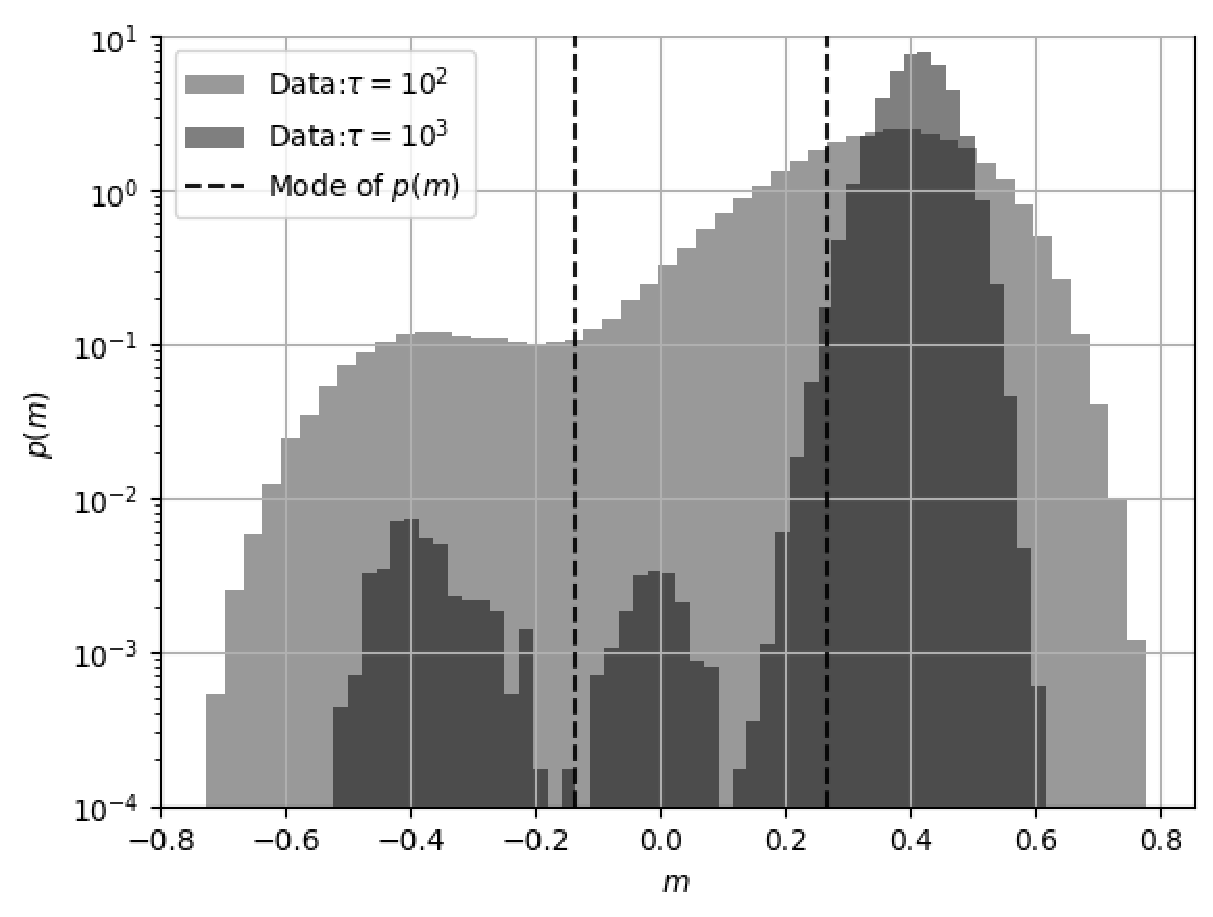} \\
\includegraphics[width=8cm]{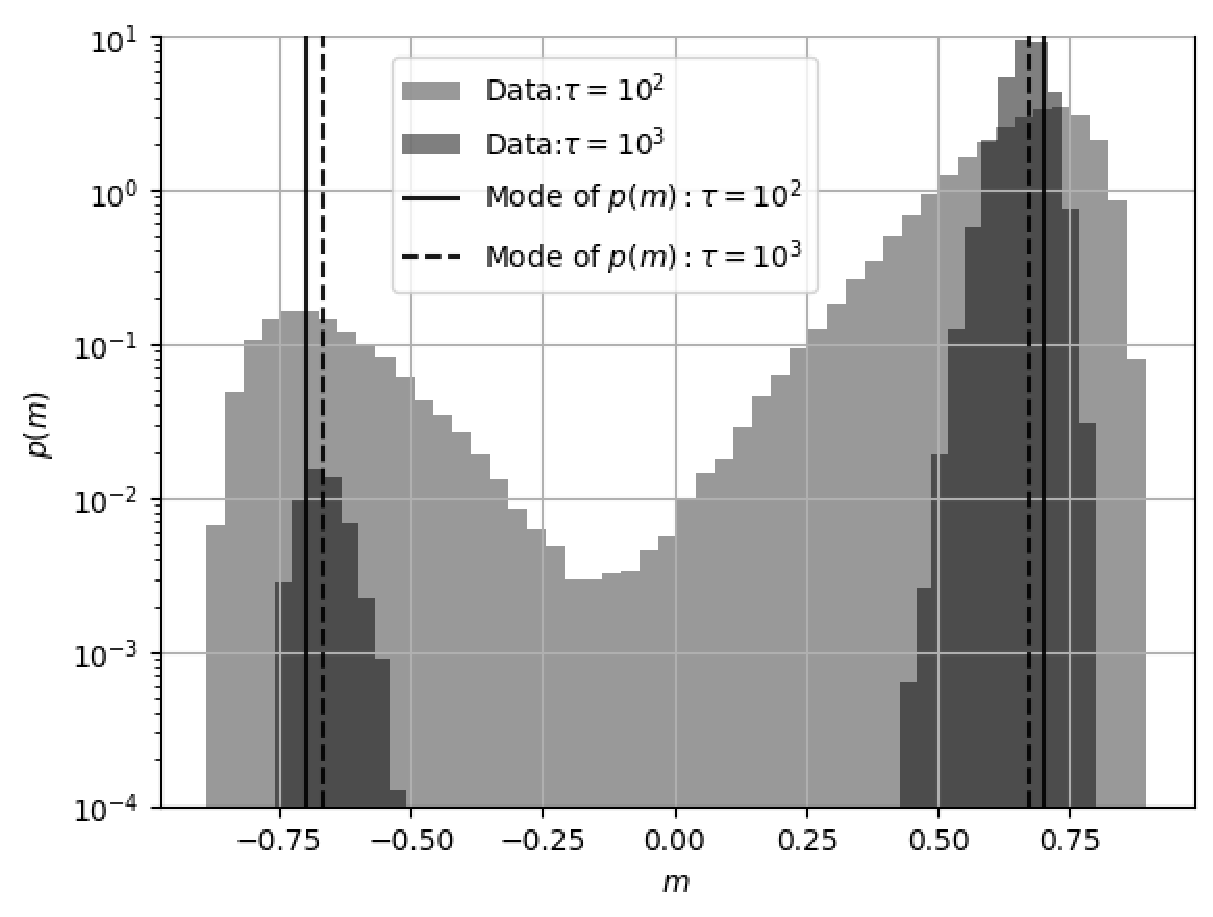}
& 
\includegraphics[width=8cm]{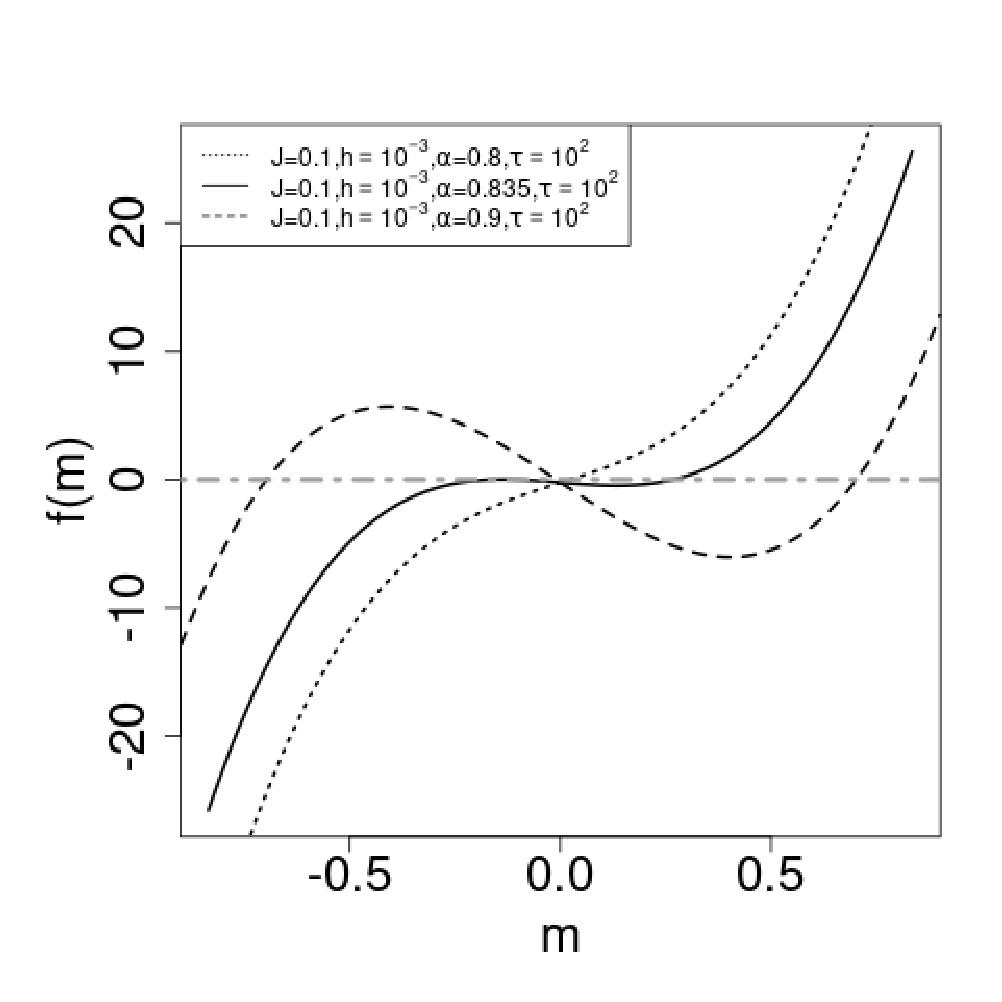} 
\end{tabular}
\end{center}
\caption{
Plot of $p(m)$ for $\alpha=0.8$ (Upper Left), $\alpha_c(h=0.001)$ (Upper Right), and $0.9$ (Lower Left),
and plot of cubic equation (\ref{eq:cubic}) vs. $m$ (Lower Right) for $J=0.1$, $h=10^{-3}$.
In the figures of $p(m)$, the gray ones show $(m)$ for $\tau=10^2$ and 
the black ones show $p(m)$ for $\tau=10^3$. The plots of the cubic equation
correspond to the cases in the three figures of $p(m)$ for $\tau=10^2$.
}
\label{fig:3}
\end{figure}

We studied the stationary distribution of $M(i,t)$. Figure \ref{fig:3} shows 
the results for the PDF $p(m)$ of $M(i,t)$. 
We adopted $J=0.1$, $h=10^{-3}$, and $\tau=10^2$ and $\tau=10^3$.
There are three figures for $\alpha=0.8$, $\alpha_c(h)$, and $\alpha=0.9$, respectively.
The fourth figure shows the plot of the cubic equation (\ref{eq:cubic})
versus $m$ for $J=0.1$, $h=10^{-3}$, and $\tau=10^2$. $\alpha_c=0.83515$ for $J=0.1,h=10^{-3},\tau=10^2$
and we chose $\alpha=0.8$, $\alpha_c$, and $0.9$. 

As one can see clearly, for $\alpha=0.8 < \alpha_c(h)$, there is a unique mode for 
$p(m)$. The variance of the PDF 
is smaller for larger $\tau$. The vertical broken line shows the position of the 
mode $m_*$ in the theory, where a discrepancy is observed.
For $\alpha=0.9 > \alpha_c(h)$, there are two modes and the values of the modes
are almost consistent with the theoretical ones. 
At $\alpha=\alpha_c(h)$, for $\tau=10^2$, the PDF has two modes, which 
is consistent with the plot of the cubic equation in the last figure (Lower Right).
The profile of the cubic equation is almost flat near $z_{t}$.
The probability current is positive for $z_{t} < z < z_{+}$, 
indicating that the lower mode should disappear finally. However, the stability
of the mode of $p_{st}(m)$ at $m = m_{t}$ is a very subtle problem.
For $\tau=10^3$, the profile of the PDF is not smooth and the result 
suggests that the equilibration is not enough for $\tau=10^3$.

\subsection{The comparison of $\alpha$-annealing with simulated annealing}
\begin{figure}[ht]
\begin{center}
\includegraphics[width=10cm]{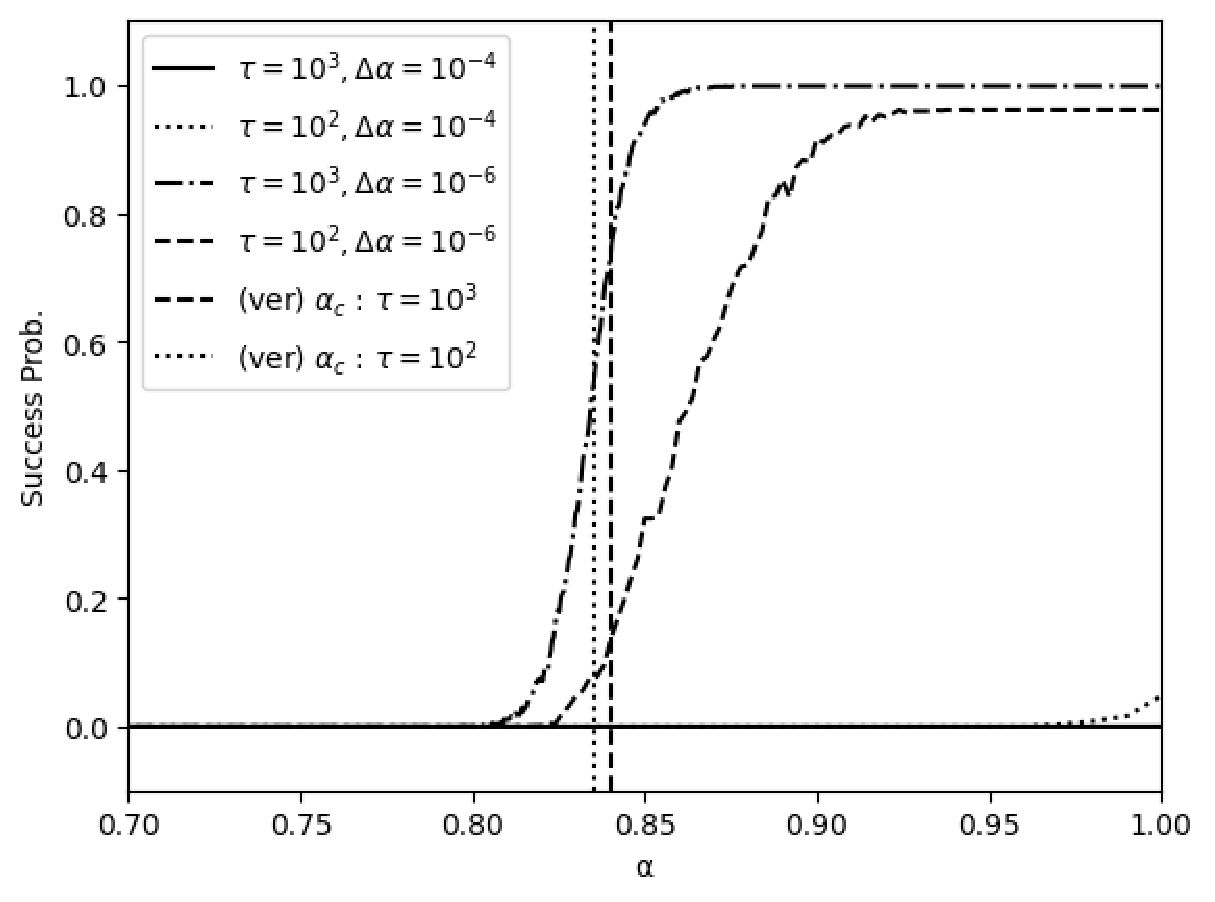}
\end{center}
\caption{Plot of $\alpha$ versus the success probabilities
to find the ground state of the Ising model.
Fast annealing processes $(T=10^4, \Delta \alpha=10^{-4})$ 
for $\tau=10^2$ (solid line) and $\tau=10^3$ (dotted line), and slow annealing processes
$(T=10^6, \Delta \alpha=10^{-6})$ for $\tau=10^2$ (broken line) and $\tau=10^3$ (chain line).
The vertical lines show 
$\alpha_c$ for $J=0.1$ and $h=10^{-3}$. }
\label{fig:4}
\end{figure}

We studied the performance of $\alpha$-annealing in ACO.
We determined $\{X(i,t)\}$ from $\{M(i,t)\}$ by $X(i,t)=\theta(M(i,t))$, 
where $\theta(x)$ is the step function,
i.e., $\theta(x)=1$ for $x>0$ and $\theta(x)=0$ for $x \le 0$.
The ACO system finds the ground state of 
the homogeneous fully connected Ising model, $\{\forall i, X(i)=1\}$,
if $\{\forall i, M(i,t)>0\}$.
We counted the number of samples where $\{\forall i, M(i,t)>0\}$ holds among $10^3$ samples
and estimated the success probability.
Figure \ref{fig:4} plots the success probability versus $\alpha$.
For $\tau=10^2$ and $\tau=10^3$, there are two results for both fast and slow annealing 
processes, respectively.

In the fast annealing cases, the success probability is almost zero and 
the system cannot find the ground state. In the slow annealing cases,
the success probability begins to increase near $\alpha_c$. It
reaches 998/1000(961/1000) at $\alpha=1-1/\tau$ for $\tau=10^3 (10^2)$. 
The success probability in SA is 0.6024 for $10^4$ trials.
The results show that the performance of $\alpha$-annealed ACO is much better than 
that of SA.

When $h=0$, the critical value of $\beta$ is $\beta_c=1/J=10$ for $J=0.1$ in SA.
$\beta(t) = (2\times \tau/T)t$ reaches $\beta_c$ at $t = 5\times 10^3$ 
for $\tau=10^3, T=10^6$.
It is a rather fast annealing process and SA cannot find the ground state with high 
success probability. When $\beta(t) \simeq \beta_c$, the correlation among the 
spin variables becomes strong and it becomes random whether $\{\forall i,X(i,t)=1\}$
 or $\{\forall i, X(i,t)=0\}$. 

As seen in Figure \ref{fig:2}, there is a continuous curve that connects the trivial 
solution $\{\forall i, M(i,t) \simeq 0\}$ for $\alpha \simeq 0$ with the correct solution 
$\{\forall i, M(i,t)>0\}$ at $\alpha=1-1/\tau$. By slow annealing of $\alpha$, 
the PDF $p(m)$ is concentrated around $m_+$ and the mode of $p(m)$ is brought along the curve.
When $\alpha(t)$ passes $\alpha_c$, the gap between $m_+$ and $m_t$ is 
large compared with the width of $p(m)$ for $\tau=10^3$.
At $\alpha=1-1/\tau$, it is possible to keep the PDF around $m_+ \simeq 1$
for $\tau=10^3$. The system can find the ground state with high success probability.
For $\tau=10^2$, the width of $p(m)$ is wide and jumps from $m_+$ to
$m_t$ occur. As a result, the success probability becomes small.
In fast annealing cases ($T=10^4, \Delta \alpha=10^{-4}$), the equilibration of 
$M(i,t)$ is not sufficient and it is difficult to align all $M(i,t)$. The success 
probability is much lower than the result of SA.

\section{Conclusion}
\label{sec:conclusion}

This paper has explored the effectiveness of $\alpha$-annealing within the Ant Colony Optimization (ACO) framework, 
particularly in seeking the ground state of the infinite-range Ising model. Our analysis, underpinned by Stochastic Differential 
Equations (SDEs), revealed that the joint probability density function (PDF) of the pheromone ratios is composed of two factors: 
entropy from the Ant System (AS) and energy from the Ising model. The parameter $\alpha$ plays a crucial role in balancing these 
factors, providing a mechanism to adjust the system's focus from broad exploratory searches to more targeted exploitative searches 
as $\alpha$ increases.

We demonstrated that a smaller $\alpha$ initially enables the system to avoid premature convergence by maintaining a broad 
exploration space, which is vital for escaping local minima. 
As $\alpha$ increases, the exploration space narrows, allowing for intensive exploration around promising regions previously identified. 
This dynamic is akin to the principles observed in quantum annealing, making $\alpha$-annealing a potent strategy for navigating complex 
optimization landscapes.

Moreover, the careful management of $\alpha$ and $\tau$—particularly the rate of pheromone evaporation—is shown to be essential for 
the system's ability to equilibrate and ultimately find the global minimum. Similar to temperature control in simulated annealing, 
$\alpha$ and $\tau$ control in $\alpha$-annealing ensures that the system can effectively balance between exploration and exploitation, 
adapting to the complexity of the optimization challenges.

In conclusion, $\alpha$-annealing emerges as a sophisticated and efficient strategy for 
enhancing ACO's performance in complex optimization scenarios. This study not only underscores 
the potential of $\alpha$-annealing as a viable alternative to traditional optimization 
techniques like simulated annealing but also highlights its unique ability to manage and manipulate exploration spaces dynamically. 
Future work will explore further applications of $\alpha$-annealing across different types of optimization problems, seeking to 
generalize these findings and refine the approach for broader practical implementation.

\subsubsection*{Acknowledgements.}
This work was supported by JPSJ KAKENHI [Grant No.{} 22K03445].

\bibliography{ref}% Produces the bibliography via BibTeX.

\end{document}